\documentclass[twocolumn,prl,sort&compress,floatfix]{revtex4}

\usepackage{amssymb}
\usepackage{amsmath}
\usepackage[dvipsnames]{xcolor}

\usepackage{hyperref}
\usepackage{graphicx}
\begin{document}

\title{Flow of deformable droplets: discontinuous shear thinning and velocity oscillations}
\author{M. Foglino$^1$, A.~N. Morozov$^1$, O. Henrich$^2$, D. Marenduzzo$^1$}
\affiliation{SUPA, School of Physics and Astronomy, University of Edinburgh, Peter Guthrie Tait Road, Edinburgh, EH9 3FD, UK \\
$^2$ SUPA, Department of Physics, University of Strathclyde, Glasgow, G4 0NG, UK}

\begin{abstract}
We study the rheology of a suspension of soft deformable droplets subjected to a pressure-driven flow. Through computer simulations, we measure the apparent viscosity as a function of droplet concentration and pressure gradient, and provide evidence of a {\it discontinuous} shear thinning behaviour, which occurs at a concentration-dependent value of the forcing. We further show that this response is associated with a nonequilibrium transition between a `hard' (or less deformable) phase, which is nearly jammed and flows very slowly, and a `soft' (or more deformable) phase, which flows much more easily. The soft phase is characterised by flow-induced time dependent shape deformations and internal currents, which are virtually absent in the hard phase. Close to the transition, we find sustained oscillations in both the droplet and fluid velocities. Polydisperse systems show similar phenomenology but with a smoother transition, and less regular oscillations. 
\end{abstract}

\maketitle

Concentrated suspensions of colloidal particles in a liquid solvent are often found in industry and nature. Familiar examples include paint, ink, food like mayonnaise and ice cream, and biological fluids such as blood~\cite{Jones}. The flow properties of colloidal suspensions can be distinctively non-trivial: for example, a suspension of colloidal spheres in water first exhibits shear thinning and then shear thickening, as the external forcing (pressure gradient or shear) is increased~\cite{Jones,jamming_rheology_Cates,Rheol_studies_2,PRL_Guy_shear_thickening,DST_Cates}. In dense suspensions, the fact that shear thickening can be {\it discontinuous} has recently attracted a lot of attention: this behaviour marks a transition between a lubrication-dominated and a frictional flow regime~\cite{DST_Cates}.

Often, in such colloidal fluids, the dispersed particles are not hard, but soft, and deformable~\cite{RBC_Pagonabarraga}. Examples are the fat droplets found in milk, or eukaryotic cells: all these can deform under flow, or when subjected to a mechanical stress. While hard sphere fluids have been studied extensively, and provide the basis for our understanding of the glass transition~\cite{Glass_transition_Cates,Glass_transition_Weeks,Glass_Pusey} and of soft glassy rheology~\cite{Glassy_material_Cates_Fielding}, less is known about the flow response of suspensions of deformable particles~\cite{RBC_Pagonabarraga,AIP_droplets,AIP_droplets_2,Droplets_shear_Loewenberg,Droplets_extensional_shear_Zinchenko,Droplets_extensional_shear_Zinchenko2}. \color{black} Nonetheless, there is a number of examples suggesting that the physics of soft suspensions is both highly interesting and important in applications. For example, experiments and simulations have recently demonstrated that glass transitions and jamming can be observed in dense monolayers of living cells\cite{Glassy_tissue_expt_manning,Glassy_cells_expt_angelini,MotilityDrive_jamming_manning,Glass_Chiang}. \color{black} Emulsions -- which are dispersion of liquid droplets in a continuous medium -- are also used in medicine and food, and their flow properties play a pivotal role in applications. Particle deformability is important to determine the rheology of a material: for example, emulsions and foams do not normally display shear thickening, unlike hard sphere colloidal fluids. 

Here, we use 2D lattice Boltzmann simulations to investigate the dynamics of a suspension of soft, and non-coalescing, droplets (Figs.~\ref{Fig_1}a,b) under pressure-driven flow within a channel. Two key parameters determine the flow response of our system: (i) the concentration, $\Phi$, defined as the ratio between the area of all droplets and the total area of the simulation domain, and (ii) the applied pressure difference $\Delta p$ driving the flow. 
We calculate the apparent viscosity, $\eta$, of the suspension as a function of $\Phi$ and $\Delta p$ -- this is the focus of the current paper, in contrast to previous numerical works on droplet rheology \cite{AIP_droplets,AIP_droplets_2,Droplets_shear_Loewenberg,Droplets_extensional_shear_Zinchenko}. \color{black} As expected, we find that, for a fixed $\Delta p$, $\eta$ increases sharply with $\Phi$, as droplets approach jamming. The behaviour of $\eta$ with $\Delta p$ is more surprising, and constitutes our central result: if the concentration is large enough, we find $\eta$ shows {\it discontinuous} shear thinning rheology. This non-Newtonian behaviour signals a nonequilibrium (flow-induced) transition between a `hard' (or less deformable) phase, where the flow does not appreciably affect droplet shape, and a `soft' (or more deformable) phase, where the droplets deform substantially, in a time-dependent fashion \color{black}. Close to the transition, we find strong and sustained oscillations in the velocity of the droplets (or of the underlying fluid). These oscillations are strikingly similar to those observed for hard colloidal particles under Poiseuille flow~\cite{velocity_oscillations}. In our case, the underlying mechanism is the proximity to the discontinuous shear-thinning transition, which leads to unabating hopping between the hard and soft viscosity branches. Discontinuous shear thinning and oscillations are more easily seen in a monodisperse suspension: in a binary system with droplets of two different sizes the transition is much smoother, more akin to a crossover.
To study the hydrodynamics of our soft droplet fluid, we follow the evolution of: (i) phase-field variables describing the density of each of the droplets, $\phi_i$, $i=1,\ldots, N$, where $N$ is the total number of droplets, and (ii) the velocity field of the underlying solvent $\mathbf{v}$. The equilibrium behaviour is governed by the following free energy density, 
\begin{equation}
\label{free_energy}
f = \frac{\alpha}{4}\sum_i^{N}\phi_i^2(\phi_i-\phi_0)^2 + \frac{K}{2}\sum_i^{N}(\nabla\phi_i)^2  + \epsilon\sum_{i,j,i<j}\phi_i\phi_j. 
\end{equation}
In Eq.~\ref{free_energy}, the first two terms ensure that each droplet is stable, as, for every $i$, $\phi_i=\phi_0$ inside the $i$-th droplet, and $0$ otherwise; these two terms also determine the surface tension of each of the droplets as $\gamma=\sqrt{(8K\alpha)/9}$ and their interfacial thickness as $\xi=5\sqrt{K/(2\alpha)}$~\cite{LB_Pagonabarraga}. The third, final term describes soft repulsion pushing droplets apart when they overlap: \color{black} $\epsilon>0$ controls the strength of this repulsion. 

The dynamics of the compositional order parameters $\{\phi_i\}_{i=1,\ldots,N}$ evolve according to a set of Cahn-Hilliard-like equations, 
\begin{equation}
\label{cahn-hilliard}
\frac{\partial \phi_i}{\partial t} + \mathbf{\nabla} \cdot (\mathbf{v}\phi_i) = M\nabla^2 \mu _i 
\end{equation}
where $M$ is the mobility and $\mu_i=\partial f/\partial \phi_i-\partial_{\alpha}f/\partial (\partial_{\alpha}\phi_i)$ is the chemical potential of the $i$-th droplet. Eq.~\ref{cahn-hilliard} conserves the area of each of the droplets (i.e., the integral of each $\phi_i$ over the whole simulation domain). 

The solvent flow obeys the Navier-Stokes equation, 
\begin{equation}
\label{navier-stokes}
\rho \Big(\frac{\partial}{\partial t} + {\mathbf v}\cdot {\nabla}\Big){\mathbf v} = -{\nabla p} - \sum_i \phi_i{\nabla}\mu_i + \eta_0\nabla^2{\mathbf v},
\end{equation}
where $\rho$ indicates the fluid density, $p$ denotes its pressure and $\eta_0$ the solvent viscosity. The term $\sum_i \phi_i{\nabla}\mu_i$ represents the internal forces due to the presence of non-trivial compositional order parameters, and as such it can also be expressed as a divergence of a stress tensor~\cite{Binary_Cates}. In what follows, we report results from hybrid lattice Boltzmann (LB) simulations~\cite{LB_method,HybridLB_Gonnella} where Eq.~\ref{navier-stokes} is solved by an LB algorithm, and Eqs.~\ref{cahn-hilliard} are solved via a finite difference. We consider flow in a channel with no-slip boundary conditions at the top and bottom walls. The flow is driven by a fixed, externally imposed pressure difference, leading to Poiseuille flow in an isotropic fluid, and neutral wetting boundary conditions for each of the droplets~\cite{Neutral_wetting_Gonnella} (see SI). Parameters used are listed in the SI, together with Reynolds and capillary numbers\color{black}, and values of $\Delta p$ in what follows are given in simulation units. While the trends we discuss are generic, the simulations we report can be mapped to a system with $\sim 100\mu$m-size droplets whose surface tension is $\gamma\sim$mN/m (see SI), embedded in a background Newtonian fluid with viscosity $\eta_0=10$ cP. Our model differs from that used in~\cite{Succi_LB} to study the glassy dynamics of foams and sprays, which in general allows for droplet coalescence.


Figure~\ref{Fig_1} shows two typical snapshots of our system, under weak pressure-driven flow and for two different values of $\Phi$. These snapshots clarify that, when $\Phi$ is low, we obtain a suspension of well-separated droplets: while these droplets interact hydrodynamically and may in principle deform, there is a substantial region between them occupied by the background solvent (Fig.~\ref{Fig_1}a). At larger concentrations, droplets touch each other even in the absence of flow, to form a percolating foam (Fig.~\ref{Fig_1}b). The snapshots also highlight that the neutral wetting boundary conditions we use lead to spreading on droplets close to the wall, with a contact angle of $90^{\circ}$. We note that droplets need to approach the wall close enough in order to stick: this only happens for above $\Phi\simeq 35\%$. 

\begin{figure*}
\centering
\includegraphics[width=0.8\textwidth]{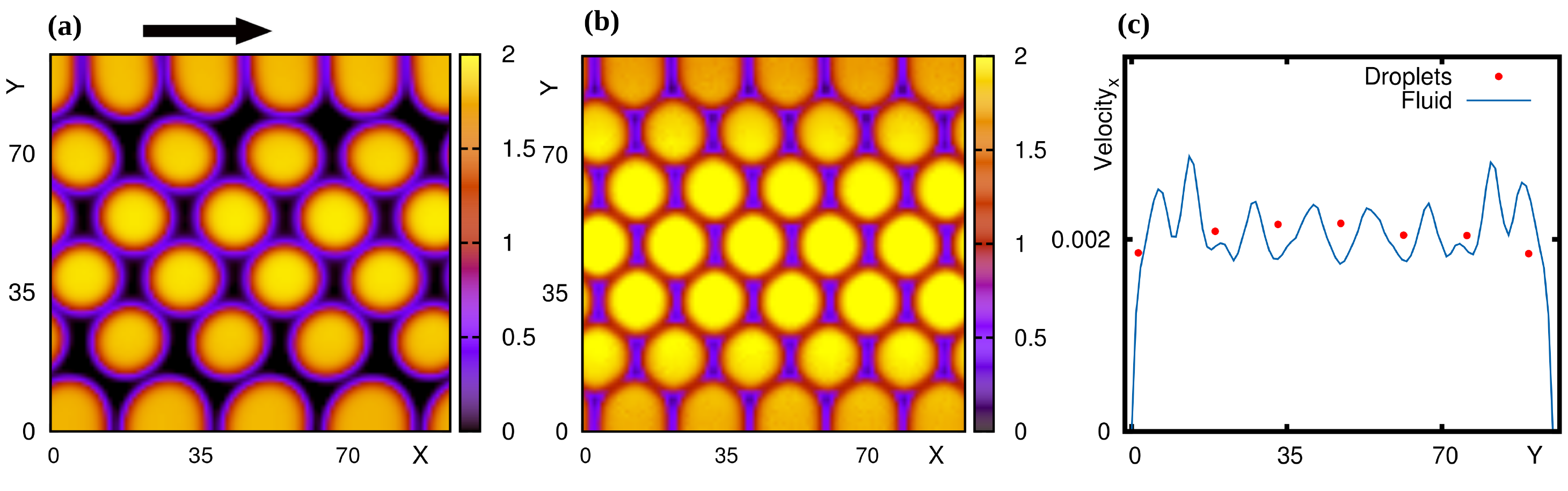}
\setcounter{figure}{0} 
\caption{Geometry and set-up. (a,b) Snapshots of a droplet suspension of area fraction $\Phi =  54.5 \%$ (a), and $\Phi = 76.3 \%$ (b). 
The color code refers to the value of $\sum_i \phi_i$: this is $\sim 2$ for droplets, and $\sim 0$ for the background solvent -- values are slightly $<2$ within boundary droplets due to spreading. (c) Velocity profile of the fluid (solid line) compared to the droplet velocity (filled circles), corresponding to a simulation with $\Phi=76.3\%$. 
}
\label{Fig_1}
\end{figure*}

We start by analysing the flow profile. To do so, we plot the average velocity along $x$ (the velocity direction) as a function of $y$ (the velocity gradient direction). While for low $\Phi$ the profile is approximately parabolic (SI, Fig.~S1), the flow becomes plug-like at higher $\Phi$ 
(see Fig.~\ref{Fig_1}c). We note that the velocity of the droplets remains close to that of the fluid throughout the channel. 

Pressure-driven flow in these suspensions is therefore strongly non-Newtonian, at least for foam-like structures with large $\Phi$.  We can nevertheless define, as in experiments, an apparent viscosity, $\eta$, by analysing the throughput flow $Q$=$\int dy v_x(y)$. A useful quantity is the ratio between $Q$ and the throughput flow of a Newtonian fluid with viscosity $\eta_0$, that of the underlying solvent (when no droplets are present). The inverse of this ratio gives a measure of $\eta/\eta_0$. A plot of $\eta/\eta_0$ as a function of $\Phi$ for a given value of $\Delta p$ (Fig.\ref{Fig_2}, inset) \color{black} shows that viscosity increases sharply and non-linearly with $\Phi$, which is suggestive of jamming as the droplet concentration increases~\cite{Jamming_Nagel,Review_Glass_Jamming_Ikeda}. Our current data, though, do not allow us to conclude whether, for small $\Delta p$, there is a linear regime with a finite albeit large viscosity at any $\Phi$, or whether $\eta \to\infty$ for $\Delta p\to 0$ above a critical $\Phi_c$, as in jammed granular flow~\cite{Jamming_Nagel,Review_Glass_Jamming_Ikeda}.\color{black}  

In Figure~\ref{Fig_2} we focus instead on the variation of $\eta/\eta_0$ with pressure difference, at fixed $\Phi$. For all concentrations, we find strong shear thinning. 
This behaviour resembles that seen in experiments probing the rheology of emulsions and foams.
Remarkably, though, for $\Phi \stackrel{\sim}{>} 50\%$ we find this shear thinning behaviour to be ``discontinuous'': in other words, there is a jump  in the viscosity for a critical value of the forcing (arrows in Fig.~\ref{Fig_2}), signalling a possible flow-induced nonequilibrium transition. This behaviour contrasts with the smooth (or continuous) shear thinning found in hard colloidal dispersions at intermediate shear rates~\cite{Jones}, or in previous simulations of droplet emulsions under shear~\cite{AIP_droplets,Droplets_shear_Loewenberg,Droplets_extensional_shear_Zinchenko}. \color{black}   
Discontinuous shear thinning is observed for $\Phi = 52.4\%, 54.5\%, 65.4\%$ and $76.3\%$, with the jump occurring for larger pressure differences as $\Phi$ increases (the trend is approximately linear, see SI, Fig.~S3\color{black}). A viscosity jump is also present for larger system size than in Figure~\ref{Fig_2} (Fig.~S4), while it is absent for non-wetting boundary conditions (Fig.~S5).\color{black} 

\begin{figure}[h!]
\centering
\includegraphics[width=.47\textwidth]{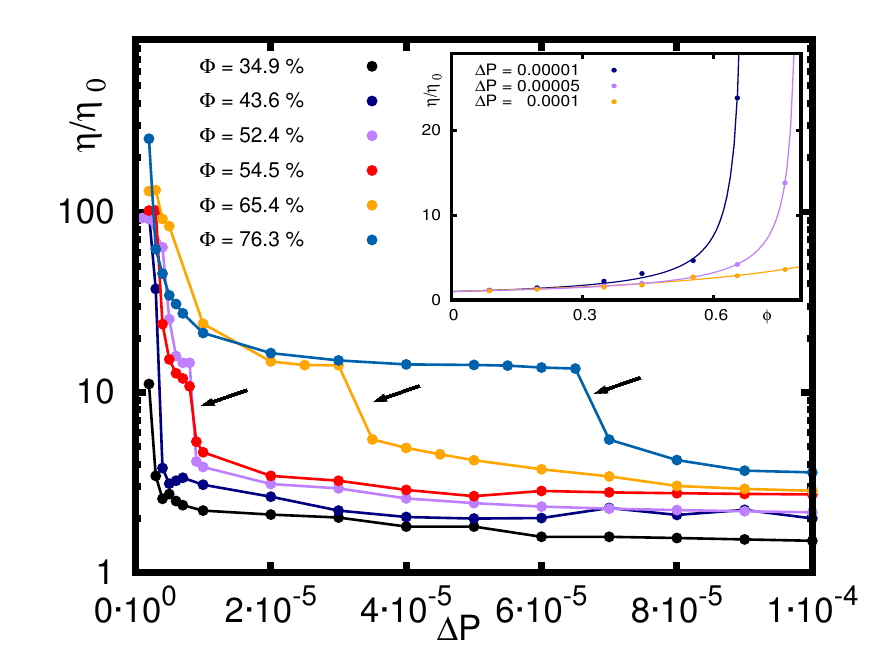}
\setcounter{figure}{1} 
\caption{Plot of the relative viscosity of the droplet suspension as a function of $\Delta p$, for five different values of $\Phi$. 
For large enough $\Phi$ this shear thinning is discontinuous,and arrows denote the disconuities. The small kinks in the bottom two curves at low $\Delta P$ are not significant, and due to inaccuracies in sampling $\eta$ by time averaging which are larger in that regime.  Inset: plot of the apparent viscosity versus $\Phi$ for three different values of $\Delta p$; fits are a guide to the eye.}
\label{Fig_2}
\end{figure}

For sufficiently large $\Phi$, therefore, we can define two viscosity branches, lying either side of the discontinuity. To identify the difference between the left and right viscosity branches, we first characterise how the flow affects droplet shape (Figs.~\ref{Fig_3}a,b). To do so, we compute the semiaxes of the ellipse defined by the inertial tensor of each droplet~\cite{Gyration} (see SI) \color{black} 
The results point to a clear difference: on the left (high viscosity) branch, the droplet shape is constant over time (Fig.~\ref{Fig_3}a, and SI \cite{SI}, Suppl. Movie 1); on the right (low viscosity) branch, there are more significant deformations, and, importantly, these display marked variations over time (Fig.~\ref{Fig_3}b, and SI, Fig.~S2 and Suppl. Movie 2)\color{black}. We therefore name the left branch `hard', and the right branch `soft'. We conclude that the discontinuity in the apparent viscosity shown in Fig.\ref{Fig_2} can be interpreted as a transition (or sharp crossover) \color{black} between a hard phase, where the droplets are effectively rigid, and a soft one, where they are highly deformable. 

\begin{figure}[h]
\centering
\includegraphics[width=.47\textwidth]{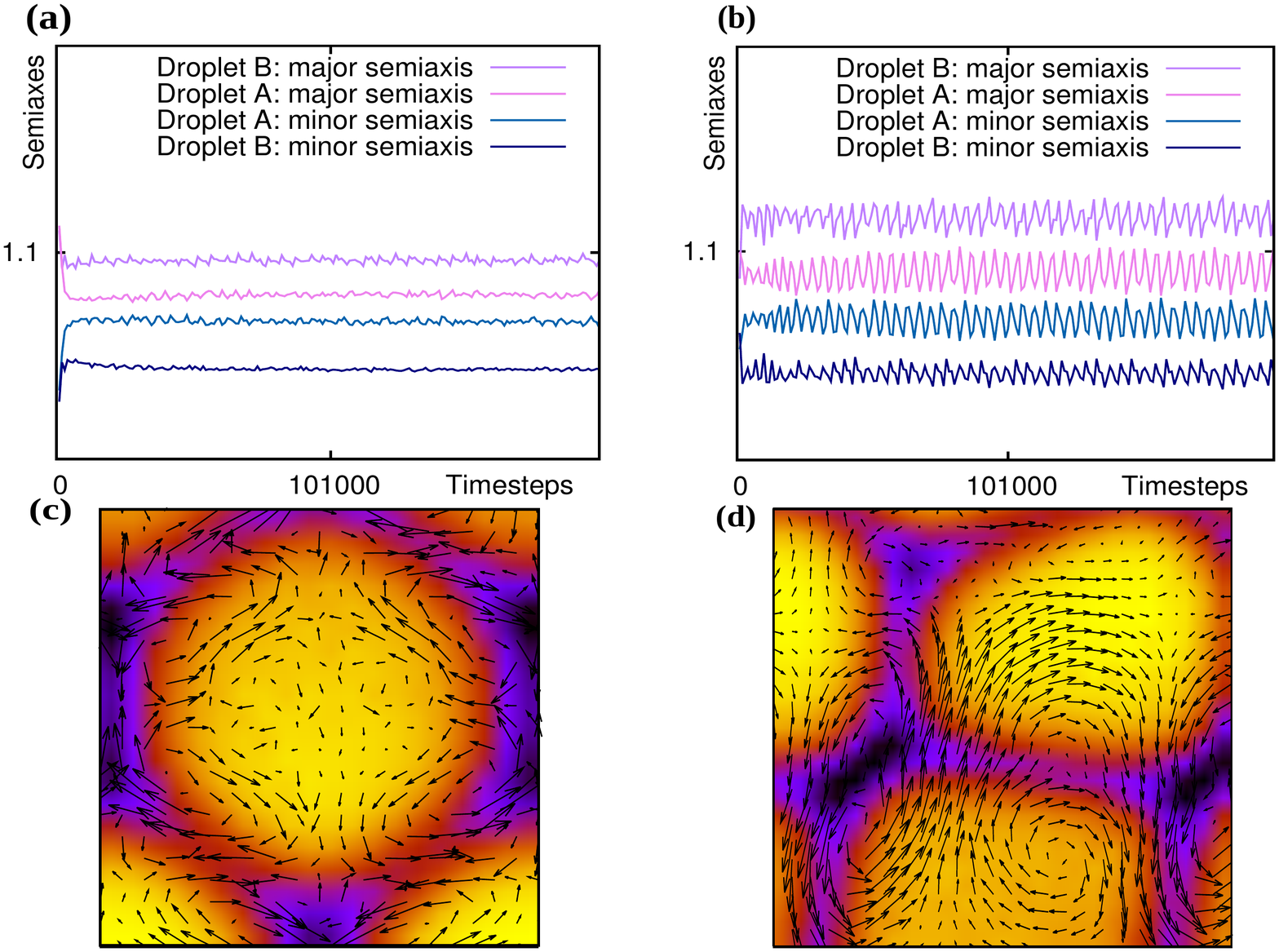}
\setcounter{figure}{2} 
\caption{(a) Plot of the two semiaxis of the ellipse defined by the inertial tensor \color{black} of two selected droplets taken from a suspension with $\Phi = 76.3 \%$ and $\Delta p = 10^{-5}$. Droplet A belongs to the central array of droplets and is far from the boundary. Droplet B belongs to the array just above the wetting layers of droplets. (b) Same as (a), but for a pressure difference of $\Delta p = 7 \times 10^{-5}$, close to the point at which the viscosity drops sharply. (c) Fluid velocity field and droplet pattern for a suspension with $\Phi = 76.3 \%$ and $\Delta p = 3 \times 10^{-6}$. (d) Same as (c), but with $\Phi = 76.3 \%$ and $\Delta p = 9\times 10^{-5}$.}
\label{Fig_3}
\end{figure}

The hard and soft branches also differ in the flow patterns observed in the steady state. When the average flow along $x$ is subtracted out, the residual flow is mainly limited \color{black}  to gaps between droplets in the hard branch (Fig.~\ref{Fig_3}c), whereas it penetrates more deeply \color{black} within the droplet interior in the soft branch (Fig.~\ref{Fig_3}d). The larger \color{black} internal flow in the soft phase arises mainly due to interaction between neighbouring lanes of droplets, and is therefore maximal close to the boundary, where the wetting layer of droplets leads to the largest effective friction with the rest of the suspension. 


Closer inspection of the simulation results reveal a further intriguing phenomenon. Consider for example a suspension with $\Phi=54.5\%$, for $\Delta p=10^{-5}$ -- i.e., just after the drop in viscosity \color{black} in Fig.~\ref{Fig_2} (the corresponding dynamics is shown in SI, Suppl. Movie 3).  As the flow sets in, the droplets that are initially close to the boundary (droplets a and e in Fig.~\ref{Fig_4}a) stick to the wall and slow down dramatically, while those close to the centre (droplets b, c, and d) undergo a `stick-slip' motion whereby they periodically accelerate and decelerate. The oscillations are visually clear when tracking droplet velocity over time (Fig.~\ref{Fig_4}a); Fourier transforming these data shows they are also very regular (Fig.~\ref{Fig_4}b). 

\begin{figure}[h]
\centering
\includegraphics[width=.49\textwidth]{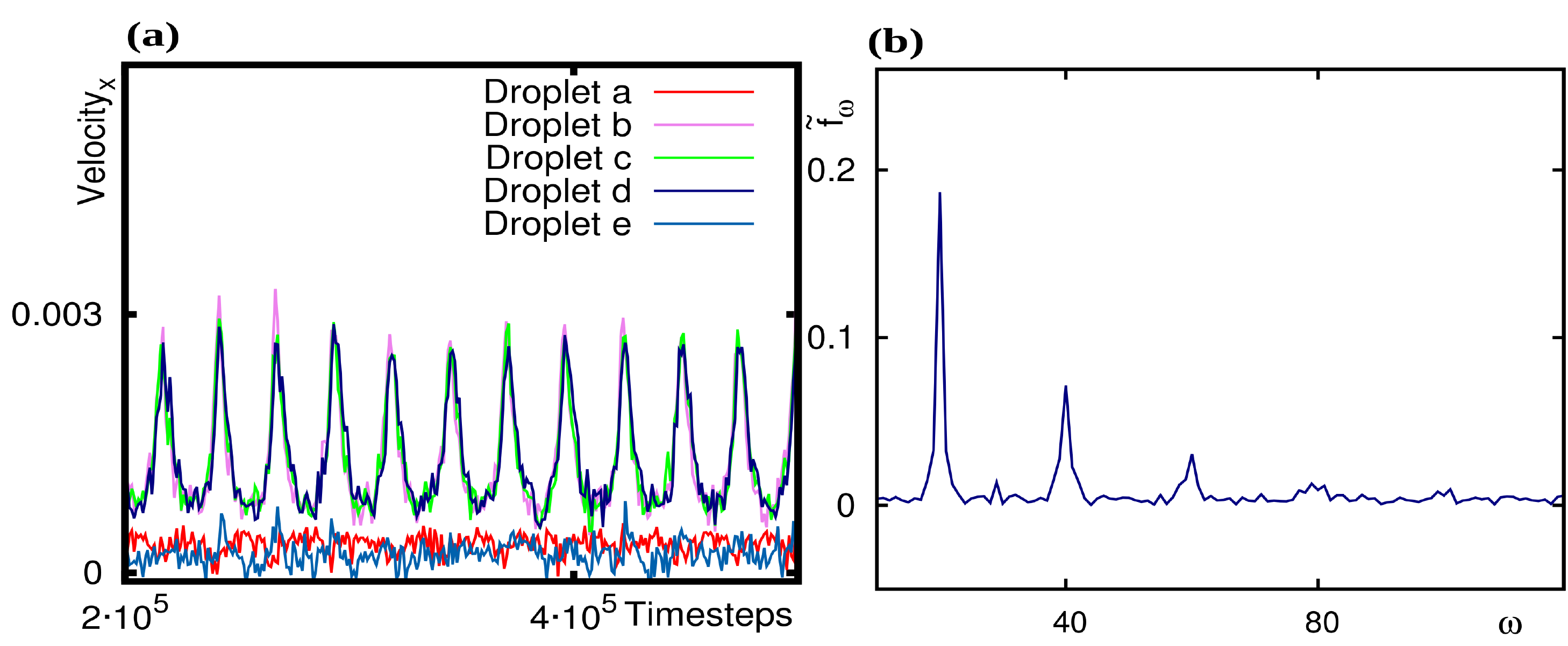}
\setcounter{figure}{3} 
\caption{\small  (a) Plots of the droplet velocities as a function of time. Droplets b, c and d belong to the array at the centre of the channel, while droplets a and e belong to the wetting layer. Oscillations in the $x$-component of the droplet velocities are apparent for all droplets except those in the wetting layer, where motion is slow. (b) Fourier transform of the droplet velocities time series: clear peaks are visible, corresponding to the oscillation frequency and its multiples. }
\label{Fig_4}
\end{figure}

Oscillations also occur for other values of $\Phi$ and $\Delta p$, and for larger system size \color{black} (see SI, Fig.~S4 \color{black} and Table S1): notably, the region in phase space where they do is, in all cases, close to the discontinuity in the viscosity curve. This is reasonable, as we expect that near the discontinuous shear thinning transition there should be hysteresis -- similarly to what happens next to a thermodynamic first-order transition. Consequently the suspension can hop between the hard and soft viscosity branches, giving rise to oscillations. An analysis of Suppl. Movie 3 additionally suggests that oscillations correlate well with deformations arising from contact interactions between the wetting layer and the nearest lane of droplets. Thus, each droplet in that array slows down when it first touches one of the droplets in the wetting layer, while it squeezes faster through the gap once it is deformed. This latter microscopic mechanism is consistent with the former explanation that oscillations require proximity to the hard-to-soft transition, because shape deformations -- which play a key role in the microscopic argument -- define the soft phase. Consistently with these arguments, we observe no transition, and no oscillations, with non-wetting boundary conditions (see SI, Suppl. Movie 4, for a typical example of flow at high $\Phi$ without droplet deformation). The droplet velocity oscillations we observe are qualitatively similar to those found in driven colloidal suspensions close to the glass transition~\cite{velocity_oscillations}, as those, too, correlate well with the gap between flowing and boundary colloids. It would be of interest to ask whether even for the colloidal case oscillations arise close to the discontinuous (shear thickening) transition. 



The suspension we have considered up to this point has been monodisperse -- all droplets had the same size. Polydispersity is known to strongly affect the behaviour of colloidal systems, for instance in so far as the glass transition is concerned~\cite{Glass_Pusey}. 
To explore the effect of polydispersity on discontinuous shear thinning in a selected case\color{black}, we show in Figure~\ref{Fig_5}a the viscosity curves of a bidisperse suspension (where the droplet radius of one component is twice as large as that of the other). While the suspension still clearly shear thins, with a comparable overall drop in viscosity with respect to the monodisperse case, here no discontinuity can be found, apart from the case of $\Phi=54.5\%$. For this concentration, we find again oscillations (Fig.~\ref{Fig_5}b), however these are much more irregular than in the monodisperse case. These findings suggest that a sufficiently strong \color{black} polydispersity smooths out the nonequilibrium transition between the hard and soft phases, turning it into a crossover. The reason is that larger droplets start to deform at a weaker forcing than smaller ones, hence the 
transition occurs more gradually with respect to the monodisperse case. 

\begin{figure}[h]
\centering
\includegraphics[width=.5\textwidth]{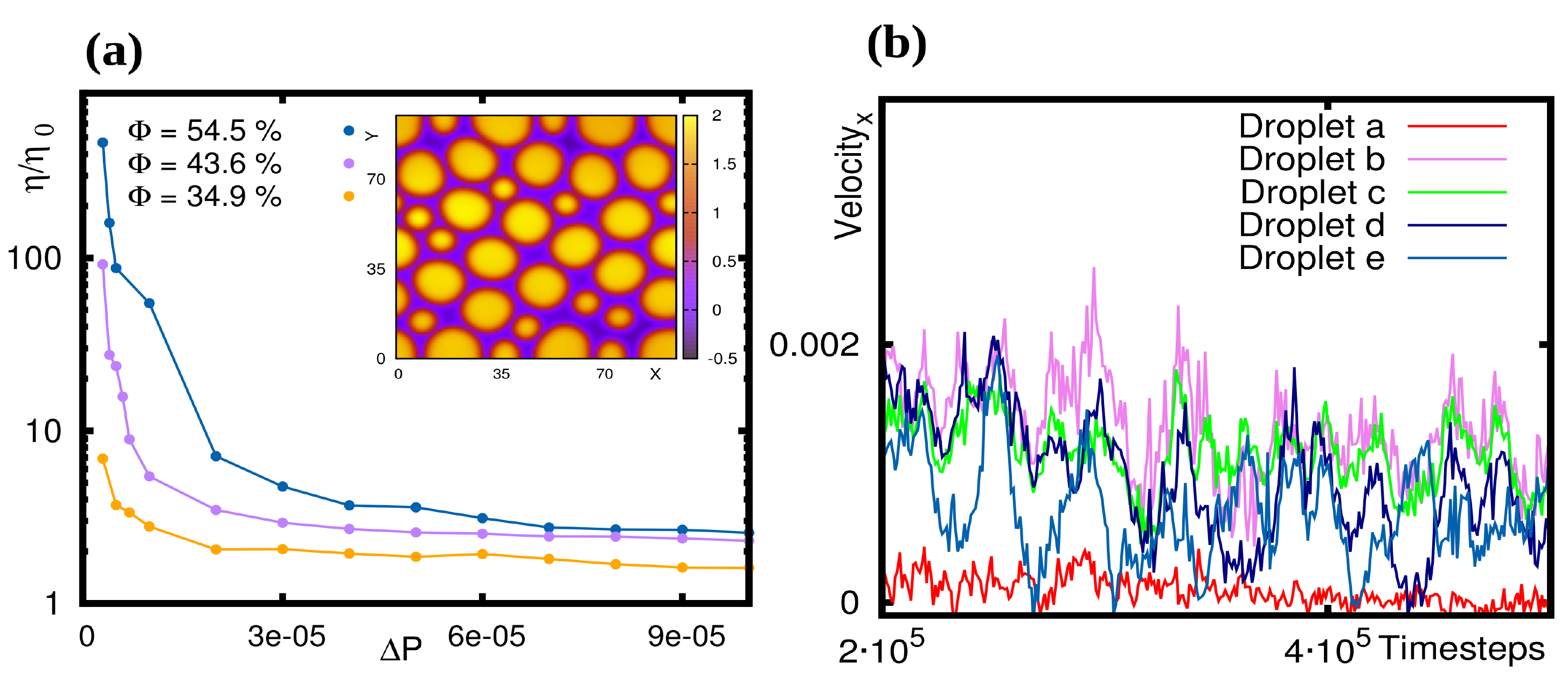}
\setcounter{figure}{4} 
\caption{Rheology of bidisperse suspensions. (a) Apparent viscosity as a function of $\Delta p$ for three different fixed values of $\Phi$, showing clear shear thinning behaviour. (b) Plot of droplets velocities versus time. Droplets b,c,d and e belong to the droplet array in the centre of the channel, while droplet a belongs to the bottom wetting layer.} 
\label{Fig_5}

\end{figure}


In summary, we have shown that the rheology of a suspension of deformable non-coalescing droplets, under a pressure-driven flow entails discontinuous shear thinning behaviour. This discontinuity may be viewed as a nonequilibrium transition between a hard droplet regime, which flows slowly, and a soft droplet phase, which flows much more readily. In the former phase, droplet shape is constant over time; in the latter, it varies significantly as they flow. \color{black} To observe discontinuous shear thicknening, we need large enough concentration, $\Phi$. At a given value of $\Phi$, our physical interpretation of the soft-to-hard transition suggests that a key dimensionless parameter is the capillary number~\cite{AIP_droplets,AIP_droplets_2,Droplets_shear_Loewenberg,Droplets_extensional_shear_Zinchenko} denoting the ratio between external forcing and interfacial tension. 
Close to the transition, we find sustained oscillations which are reminiscent of those reported previously for hard colloidal systems close to the glass transition~\cite{velocity_oscillations}. It is tempting to speculate that in both cases oscillations arise due to proximity to a discontinuous transition. 

In the future, it would be interesting to recreate discontinuous shear thinning in the lab, by studying the rheology of suspensions of non-coalescing droplets~\cite{Pastes_Cloitre}. Theoretically, our findings prompt new questions. For instance, it would be useful to characterise the dependence of the hard-to-soft transition on surface tension. It would also be informative to study the microrheology~\cite{Microrheology_Cloitre} of our droplet suspensions and see what signatures discontinuous shear thinning leaves there. 

\paragraph{Acknowledgements:} We thank ERC (COLLDENSE network) and EPSRC (EP/N019180/1) for support.

\bibliographystyle{unsrt}
\bibliography{bibliography}  
\renewcommand{\thefigure}{S\arabic{figure}}

\newpage

\end{document}


\section*{Supplementary Information}

\paragraph{Additional results} 

In this Section we provide some additional results which are referred to in the main text.

As we mentioned in the main text while discussing Figure 1c, the flow profile in our droplet suspensions depends strongly on density, $\Phi$. In Figure~\ref{Fig_S1} we plot the $\hat{x}$ component of the average velocity of the fluid as a function of $y$. As stated in the main paper, this Figure shows that for low density the profile is approximately parabolic, while the flow becomes plug-like at higher values of area fraction $\Phi$.

\begin{figure}[h!]
\centering
\includegraphics[width=.48\textwidth]{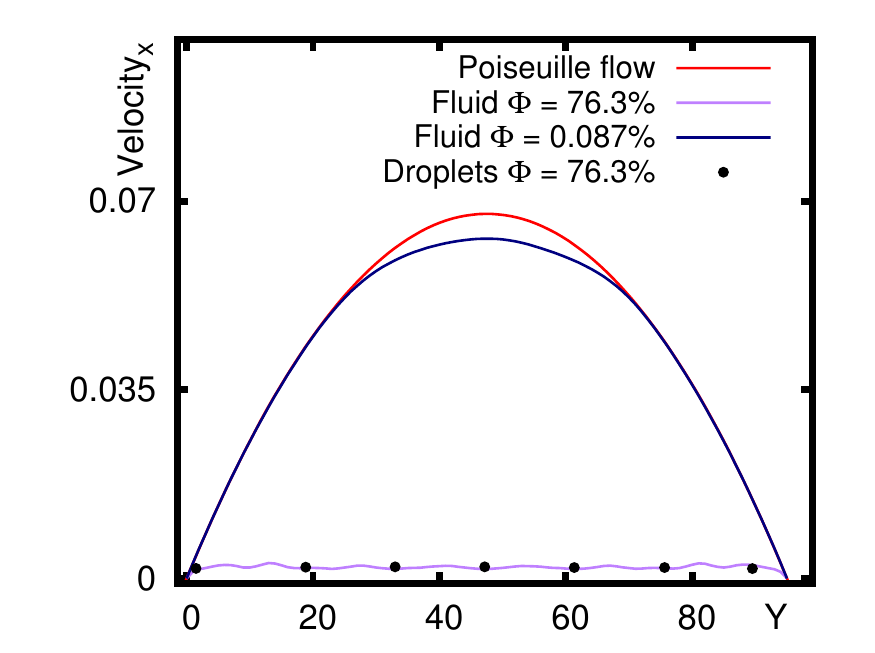}
\setcounter{figure}{0} 
\caption{Fluid velocity profile (as a function of $y$) for two different values of area fraction $\Phi$. For lower density (blue line) we observe a parabolic profile, compatible with the behaviour expected of an isotropic fluid with the same viscosity (red line), for higher density (purple line) we observe a plug-like flow. In the latter case the velocity of the droplets (solid circles) is similar to the fluid velocity (purple line).}
\label{Fig_S1}
\end{figure}

{From the velocity profile, we can define an apparent viscosity by comparing the throughput flow of our droplet suspension with that of a Newtonian fluid with viscosity $\eta_0$ (see text). The apparent viscosity increases sharply with $\phi$, as shown in the inset of Figure 2 in the main text - the fits are performed using the functional form $f(\phi)=a\exp(b/(c-\phi))$, inspired by previous work on hard colloidal suspensions~\cite{colloids_2}, where $a, b, c$ are fitting constants. These fits are only meant as guide for the eye, as our data do not allow us to discriminate between two scenarios: one in which there is a yield stress as in standard theories of jamming, and another one where there is a very large but finite linear viscosity.}

In order to give a quantitative measure of the deformability of our droplets, we construct the inertia tensor of each of the droplets~\cite{Gyration}. The square root of the larger and smaller eigenvalues of this tensor are proportional to the major and minor semiaxes of an ellipse which best approximates the droplet (the proportionality constant is $2/M$, with $M$ the ``mass'' of the droplet, equal to its area in our case). Whilst for large $\Phi$ the droplets may not well approximated by ellipses, the major and minor semiaxis sizes $\lambda_1$ and $\lambda_2$ still give a measurement of droplet anisotropy, and allow us to estimate variation of droplet shape over time. More sophisticated measures would involve higher-rank deformation tensors as in~\cite{Droplet_Yoshinaga}, but they are not necessary for our purposes here.
The droplet shape evolution marks a transition between two different behaviours, as shown in Figures 3a and 3b in the main text. As discussed in the main text, our results point to a transition between a hard branch, where the droplet shape is approximately constant in time (Fig.~3a), and a soft branch, where the droplets are subjected to significant deformation over time (Fig.~3b). More systematically, we have computed how the spread in the variation of $\sqrt{\lambda_1/\lambda_2}$ over time depends on $\Delta p$. Data (averaged over the array of droplets immediately above the wetting layer) are shown in Figure~\ref{Fig_S2}: they show a clear jump in the spread at the value of $\Delta p$ for which we observe the viscosity jump in Figure 2.

\begin{figure}[h!]
\centering
\includegraphics[width=.48\textwidth]{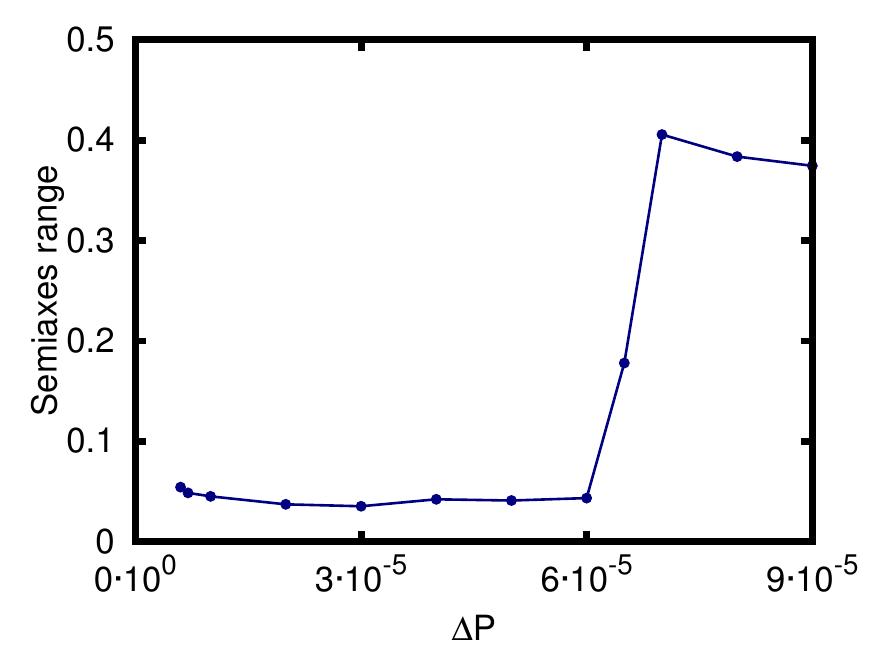}
\setcounter{figure}{1} 
\caption{Plot of the maximal temporal spread of the ratio $\sqrt{\lambda_1/\lambda_2}$, as a function of $\Delta P$. Here, $\lambda_1$ and $\lambda_2$ are the larger and smaller eigenvalue of the inertial tensor, respectively (this ratio may also be thought of the ratio between the major and minor semiaxis of an ellipse which fits the droplet shape). This plot is an average over droplet in the array next to the wetting layer, and it gives a measure of the variation of droplet shape over time -- it can be seen that such variation increases sharply where we observe discontinuous shear thinning (compare this plot with Fig. 2 in the main text).}
\label{Fig_S2}
\end{figure}

This hard-to-soft transition can also be appreciated by comparing the droplets suspension behaviour shown in Suppl. Movies 1 and 2, which refer to a suspension with $\Phi = 65.4\%$, subjected to two different values of $\Delta p$. Suppl. Movie 1 ($\Delta p = 3\cdot10^{-5}$) shows an example of flow within the hard phase: after a quick initial rearrangement of the positions of the droplets, their shape is maintained constant over the entire simulation. On the other hand, Suppl. Movie 2 shows an example of rheology in the soft phase, with the droplet shapes now undergoing clear flow-induced deformations over time. In particular, if we focus on the second array of droplets (from the bottom), it can be seen that each droplet undergoes a periodic cycle of deformations as it slides over the array of droplets stuck to the wall.

Suppl. Movie 3 shows an example of velocity oscillations, referring to a suspension of area fraction $\Phi = 54.5\%$ subjected to an imposed pressure difference $\Delta p = 10^{-5}$. After some short transient dynamics, the droplets in the suspension rearrange to form six arrays, flowing at different velocities. In particular the four arrays in the centre move as a plug, and display clearly visible velocity oscillations. Such oscillation appear consistent with the deformation sustained by the droplets when sliding over the wetting layers at the boundaries. As the droplets get stuck to the wall, they move slower because of the neutral wetting boundary condition, the ones in the neighbouring lane need to slide over and surpass them. In doing so, they periodically fall into the gaps between two successive boundary droplets, hence they are subjected to periodical deformation (Suppl. Movie 3). 
\vspace{20pt}

A plot of the location of the critical $\Delta P$ corresponding to the discontinuity in the relative viscosity curve, versus the packing fraction of the suspension is given in Fig.~\ref{Fig_S3}.

\begin{figure}[h!]
\centering
\includegraphics[width=.48\textwidth]{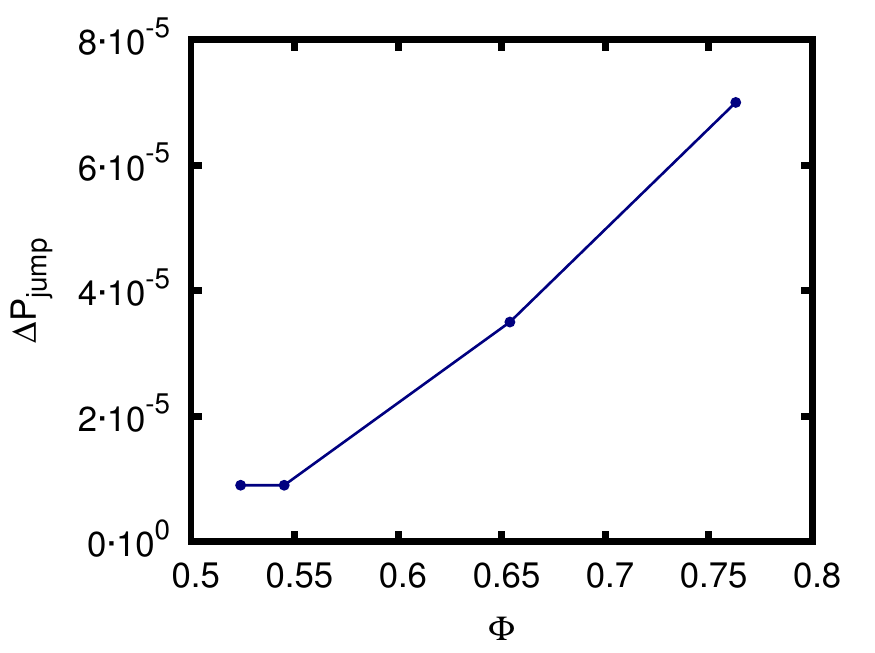}
\setcounter{figure}{2} 
\caption{Plot of the value of the applied pressure difference $\Delta P$ corresponding to the jump in the system viscosity as a function of the suspension area fraction $\Phi$. } 
\label{Fig_S3}
\end{figure}

The results in the text and thus far pertain to a system size of $L_y=96$ in the velocity gradient ($y$) direction. To address the dependence of discontinuous shear thinning on system size, we ran additional simulations of our droplet suspension confined in a channel of larger width. In particular we simulated a channel with $L_y = 134$ and analysed the flow properties of our suspension for two fixed values of area fraction, $\Phi = 54.7\% $ and $\Phi = 62.5\%$ (these cases result in two more droplet arrays than considered in the text). As we can see in Figure~\ref{Fig_S4}, also in this case we can detect the discontinuous jump in the suspension relative viscosity as a function of the applied pressure difference.	
Additionally, if we focus on the values of applied pressure difference $ \Delta P $ close to the viscosity jump, we observe oscillations in the droplets velocities over time evolution as in the previous case	of a thinner channel (see inset Fig.~\ref{Fig_S4}). The trends we report in the main text are therefore robust with changes in the system size, at least within the range we have analysed.
	
\begin{figure}[h!]
\centering
\includegraphics[width=.48\textwidth]{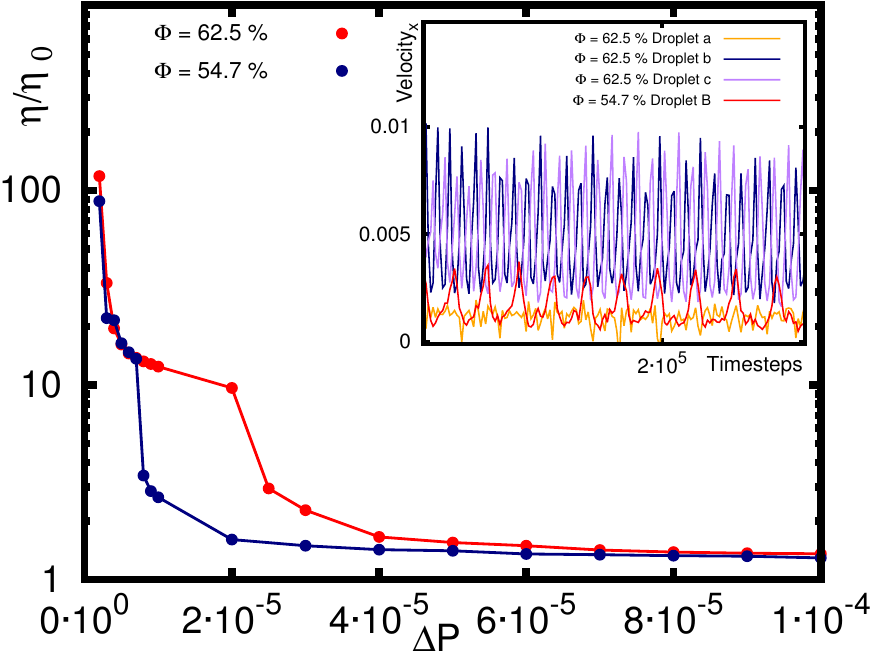}
\setcounter{figure}{3} 
\caption{Plot of the apparent viscosity as a function of $\Delta P$ for two fixed values of area fraction $\Phi = 62.5\%$ and $\Phi = 54.7\%$ for channel of width $L_y = 134$. As in the previous case of a thinner channel ($L_y=96$, see main text), we observe discontinuous shear thinning in the viscosity curves. Inset: plot of the droplets velocities oscillations over time evolution for a fixed value of applied pressure difference $\Delta P$ close to the discontinuity in the system viscosity. Droplets $b$,$c$ (for the case of $\Phi = 62.5\%$) and $B$ (for the case of $\Phi = 54.7\%$) belong to the lines in the centre of the suspension, while droplet $a$ is in contact with the wetting layer of droplets, which considerably reduce its velocity.} 
\label{Fig_S4}
\end{figure}
	
All results presented up to now refer to neutral wetting boundary condition for the compositional order parameters, $\phi_i$, at the wall. 
If, instead, non-wetting boundary conditions are used, we observe a much lower apparent viscosity with respect to the case of neutral wetting, and we find essentially no discontinuous shear thinning (Fig.~\ref{Fig_S5}). Additionally, there are no velocity oscillations in the parameter range we considered. Suppl. Movie 4 shows an example of the dynamics with a suspension with non-wetting boundary conditions. In line with the results in Figure~\ref{Fig_S5}, it can also be seen that the droplets flow without appreciable deformation. This confirms that friction with the wetting layers of droplets is needed both for the discontinuous shear thinning behaviour, and the presence of velocity oscillations. 

\begin{figure}[h!]
\centering
\includegraphics[width=.48\textwidth]{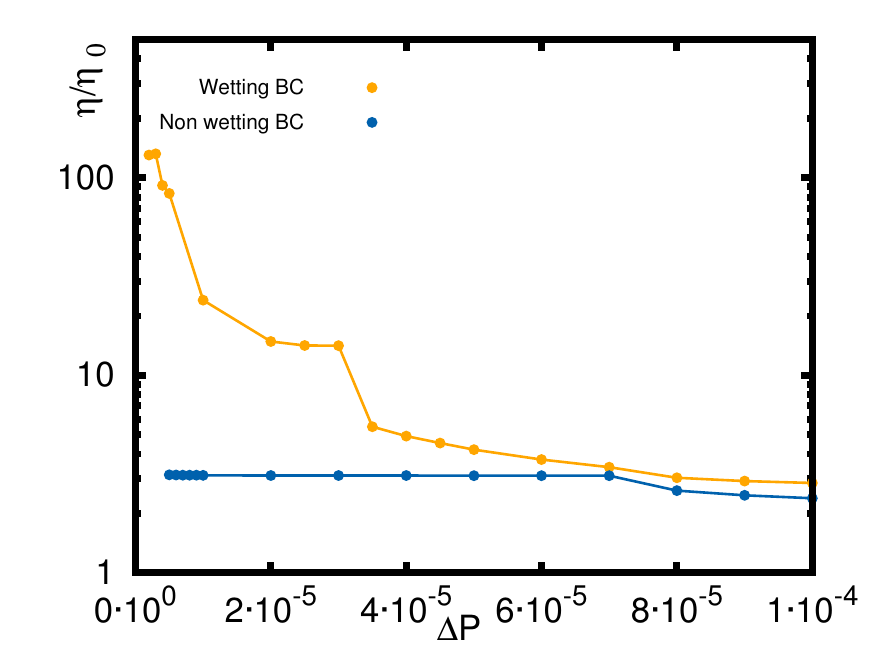}
\setcounter{figure}{4} 
\caption{Plot of the apparent viscosity as a function of $\Delta p$ for neutral wetting (top curve) and non-wetting (bottom curve) boundary conditions, for a suspension with $\Phi=65.4\%$.} 
\label{Fig_S5}
\end{figure}

\begin{table}[h!]
\begin{tabular}{ | p{3cm} | p{4cm} | } 
 \hline
 \centering Area fraction $\Phi$ & \qquad \quad Bodyforce $F (\times 10^{-5}) $  \\ [0.5ex]  
 \hline\hline
 \centering $54.5 \%$ & $1,  2,  3, 4, 5, 6, 7, 8 , 9, 10$  \\ 
 \hline
 \centering $65.4\%$ & $ 3, 4, 5, 6, 7, 8, 9, 10$    \\
 \hline
 \centering $76.3\%$ & $ 7, 8, 9, 10$  \\
 \hline
 \end{tabular}
 \label{table1}
\caption{Parameter values (simulation units) at which oscillations in the droplet velocities are found.}
\end{table}

\paragraph{Simulation details}

Here we provide more details on our simulations. 

We used a hybrid lattice Boltzmann (LB) algorithm, analogous to the one described for a binary fluid in~\cite{HybridLB_Gonnella}. Thus, the equations for the compositional order parametrs, $\phi_i$, $i=1,\ldots N$ (recall $N$ is the number of particles), were solved by means of a finite difference algorithm, whereas the Navier-Stokes equation was solved by LB (predictor-corrector algorithm, as in~\cite{LB_method_2}). 

In order to simulate a pressure gradient, a body force (force per unit density) was included in our LB algorithm, and added to the collision operator at each lattice node. The term $-\sum_i \phi_i{\nabla}\mu_i$ was also added as a body force, as this limits spurious LB velocities in equilibrium~\cite{SpuriousVelocities_Wagner}.

Neutral wetting boundary conditions were enforced by requiring that, close to one of the two flat walls sandwiching our system, the following conditions were met, for each $i=1,\ldots N$ (recall $N$ is the number of particles)
\begin{eqnarray}\label{BCs}
\frac{\partial \mu_i}{\partial z} & = & 0 \\ \nonumber
\frac{\partial \nabla^2 \phi_i}{\partial z} & = & 0.
\end{eqnarray}
In Eq.~\ref{BCs}, the first line ensures density conservation, the second determines the wetting to be neutral. Non-wetting boundaries are implemented by substituting the second line of Eq.~\ref{BCs} with the condition $\phi_i=0$ (in practice, this is enforced at the mid-point between the first and second lattice node along $y$).

Finally, we list here parameters used in this study. We fixed the droplet radius to $R=8$ ($R$ is the radius of a circle with the same area as one of the droplet) and the mobility to $M=0.1$, while free energy parameters were $\alpha=0.07$, $K=0.14$, $\epsilon=0.05$. {The viscosity of the background fluid was $\eta_0=5/3$ -- for simplicity this is also the viscosity of the fluid inside the droplet (it would be possible in principle to relax this approximation in the future, for instance by letting $\eta_0$ depend on $\phi$~\cite{adriano}).}
Simulations were run on a $96\times 96$ lattice, with periodic boundary conditions along the $x$ axis, and boundary walls along the $y$ axis while $\Phi$ and $\Delta p$ were varied as described in the main text. While the trends we have shown in the main text are generic and do not rely on a particular choice of physical units, the parameters listed above can be mapped onto a physical system with droplets of size (diameter) equal to $100$ $\mu$m, in a background fluid with viscosity $10$ cP, or $10^{-2}$ Pa s (in the absence of droplets), and where the surface tension $\gamma=\sqrt{(8K\alpha)/(9)}\sim 0.09$ corresponds to $1$ mN/m. With this mapping, a velocity of $10^{-3}$ in simulation units corresponds to $1$ mm/s. The Reynolds number in the simulations reported in Figure 2 (main text) range from $\sim 0.4$ ($\Delta P=5 \times 10^{-6}$, resulting in a fluid maximum velocity $v_{\rm max}\simeq 0.003$) to $\sim 8$ (for $\Delta P=10^{-4}$, $v_{\rm max}\sim 0.07$). These values are always small enough that no inertial instabilities are observed. The conventional capillary number, defined as ${\rm Ca} = \frac{\eta_0 v}{\gamma}$ (with $\eta_0$ solvent viscosity, and $v=v_{\rm max}$) ranges from $\sim 0.06$ (for $\Delta P=5\times 10^{-6}$) to $\sim 1.23$ (for $\Delta P=10^{-4}$).

\bibliographystyle{unsrt}
\bibliography{bibliography}